\documentclass{jfm}

\usepackage{graphicx}
\usepackage[dvipsnames]{xcolor}
\usepackage{newtxtext}
\usepackage{newtxmath}
\usepackage{natbib}
\usepackage{hyperref}
\usepackage{physics}
\usepackage[retainorgcmds]{IEEEtrantools}
\usepackage{tikz}
\hypersetup{
    colorlinks = true,
    urlcolor   = blue,
    citecolor  = black,
}

\newcommand{\RomanNumeralCaps}[1]
\linenumbers

\newcommand{\beq}{\begin{equation}}
\newcommand{\eeq}{\end{equation}}
\newcommand{\bieee}{\begin{IEEEeqnarray}{rCl}}
\newcommand{\eieee}{\end{IEEEeqnarray}}

\title{Space-time proper orthogonal decomposition of actuation transients: plasma-controlled jet flow}

\author{Brandon Yeung\aff{1}
 \and Oliver T. Schmidt\aff{1}
  \corresp{\email{oschmidt@ucsd.edu}}}

\affiliation{\aff{1}Department of Mechanical and Aerospace Engineering, University of California San Diego, CA 92093, USA}

\begin{document}
\maketitle

\begin{abstract}
We investigate the forcing-induced transient between statistically stationary and cyclostationary states. The transient dynamics of a turbulent supersonic twin-rectangular jet flow, forced symmetrically at a Strouhal number of 0.9, are studied using synchronized large-eddy simulations (LES) and space-time proper orthogonal decomposition (space-time POD).
Under plasma-actuated control, the statistically stationary jet evolves towards a cyclostationary state over a transient phase. Forcing-induced perturbations of the natural jet are extracted using synchronized simulations of the natural and forced jets. A database is collected that captures an ensemble of realizations of the perturbations within the initial transient. The spatiotemporal dynamics and statistics of the transient are analyzed using space-time POD for each symmetry component. The eigenvalue spectra unveil low-rank dynamics in the symmetric component. The spatial and temporal structures of the leading modes indicate that the initial pulse of the actuators produces large, impulsive perturbations to the flow field. The symmetric mode reveals the contraction of the shock cells due to the forcing, and shows the evolution of the mean flow deformation transient. 


\end{abstract}


\section{Introduction}
Reducing jet noise is one of the key challenges in civil and military aviation. Military-style supersonic jets, in particular, can emit intense noise, exposure to which jeopardizes the health of personnel working in close proximity. 
Whereas the accurate prediction of jet noise is increasingly attainable \citep{BresLele2019RSTA}, controlling jet noise remains more challenging. Developing effective control strategies that achieve robust noise reduction requires fundamental insights into the coherent structures in the jet that are the root cause of noise \citep{JordanColonius2013AnnuRev}. While significant inroads have been made into understanding the occurrence of such structures and the underlying hydrodynamic instabilities in high-Reynolds number, axisymmetric jets \citep{SchmidtEtAl2018JFM}, such progress has so far eluded the analysis of jets from complex nozzle geometries that are typical of military-style aircraft. The focus of this work is the supersonic twin-rectangular jet flow recently investigated by \citet{SamimyEtAl2023JFM} at Ohio State University (OSU). Their experimental efforts have included the use of localized arc filament plasma actuators (LAFPA; \citep{SamimyEtAl2004ExpFluids}) for control, and have resulted in good control authority on twin-rectangular jet flows. In parallel, we have recently conducted large-eddy simulations (LES) of the same jet \citep{YeungEtAl2022AIAA}, with plasma actuation modeled numerically \citep{YeungSchmidt2023AIAA}. Whereas past efforts at jet noise control focused on periodic forcing, transient forcing could potentially enable new actuation strategies for noise mitigation. As a first step towards future control design, in this first of its kind study, we focus on the physics of actuation transients in the plasma-controlled twin jet, with an emphasis on the deformation of its mean flow due to forcing.

When subjected to exogenous forcing, the twin-rectangular jet evolves from a wide-sense stationary state to a wide-sense cyclostationary state. In between, the jet flow satisfies neither statistical assumption. To study this actuation transient, we leverage the modal decomposition technique called space-time proper orthogonal decomposition (space-time POD; \citep{SchmidtSchmid2019JFM}). Space-time POD is the most general form of POD.
It extracts structures that are coherent in space and over a finite time horizon, and which optimally represent the second-order statistics. Since space-time POD makes no assumptions about the temporal dynamics, it is well-suited to the analysis of turbulent flows that exhibit non-ergodicity \citep{SchmidtSchmid2019JFM,FrameTowne2023PLoS}, such as a transient. More specialized forms of space-time POD have also been developed. Spectral POD (SPOD; \citep{TowneEtAl2018JFM})
requires the flow to be wide-sense stationary. Analogously, cyclostationary SPOD (CS-SPOD; \citep{HeidtColonius2024JFM}) requires the flow to be wide-sense cyclostationary. When these more restrictive conditions are met, the corresponding tools should be exploited. However, the actuation transient meets neither condition; as a consequence, only the general space-time form of POD is applicable.

In addition to space-time POD, a number of data-driven methods have been proposed to obtain modal decompositions of transient flows. Here, we list only a small sampling, with the aim of providing a flavor of the breadth of decompositions available. Arguably the most prominent, dynamic mode decomposition (DMD; \citep{Schmid2010JFM}) and its variants
have been extensively applied to transient phenomena. At their core, these methods assume exponential and oscillatory temporal dynamics, with various refinements to account for transient behaviors.
Empirical mode decomposition (EMD; \citep{HuangEtAl1998RSPA}), designed for transient signals, decomposes data into intrinsic mode functions via the Hilbert-Huang transform. 
Space-only POD has itself been modified for transient flow analysis, including the addition of shift modes \citep{NoackEtAl2003JFM}
and multi-scale POD \citep{MendezEtAl2019JFM}.
A canonical and often-studied example is the onset of vortex shedding in a laminar cylinder wake in unstable equilibrium. For the high-Reynolds number, highly nonlinear, and fully turbulent twin jet, space-time POD remains the most suitable statistical method because of its guaranteed optimality (in an energy norm) as well as its generality, since it does not assume an \textit{ansatz} for either the spatial or temporal evolution of the flow.

To better capture the effects of plasma actuation on the unsteady turbulent flow field, we employ the technique of synchronized simulations proposed by \citet{Nikitin2008JFM}
and independently by \citet{UnnikrishnanGaitonde2016JCP}. A pair of simulations, with and without actuation, are simultaneously advanced from the same initial condition. 
In this work, we use synchronized simulations to generate an ensemble of realizations of the deviation between them, then analyze the turbulent statistics of the deviation with space-time POD.

The remainder of this paper is organized as follows. Sections \ref{sec:syncLES} and \ref{sec:STPOD} recapitulate the methodologies of synchronized simulations and space-time POD, respectively. The setup of the LES is summarized in \S\ref{sec:numericalsetup}. Section \ref{sec:d2sym} describes the statistical symmetries of the twin jet. The modeling of plasma actuation is outlined in \S\ref{sec:actuation}. Results from space-time POD analysis are reported in \S\ref{sec:results}, then discussed and summarized in \S\ref{sec:discussion}.

\section{Spatiotemporal statistics of forced transient flows}
\subsection{Synchronized simulations}\label{sec:syncLES}
For detailed descriptions of the synchronized simulations technique, we refer the reader to \citet{Nikitin2008JFM}
and \citet{UnnikrishnanGaitonde2016JCP}. Here, we will only recapitulate the basics.
To study the effects of actuation on the statistically stationary but unsteady natural jet, we decompose the solution to the forced compressible Navier-Stokes equations (NSE) as $\vb*q_\mathrm{f}(\vb*x,t) = \vb*q_\mathrm{n}(\vb*x,t) + \vb*q(\vb*x,t)$,
where $\vb*q_\mathrm{n}(\vb*x,t)$ is the solution to the natural, unforced NSE and interpreted as an unsteady base state, and $\vb*q(\vb*x,t)$ is the perturbation from that base state. 
Substituting this decomposition into the forced NSE in perturbation form, then removing the unforced NSE, yields the governing equations of the perturbation.
In practice, 
an unforced LES and a forced LES are advanced in parallel from the same initial condition. At each time step, the difference between the two flow states is the solution, $\vb*q(\vb*x,t)$. This by no means amounts to a linearization: we are merely interested in the turbulent statistics of the deviation between the two transients, starting from the same initial condition.

\subsection{Space-time POD}\label{sec:STPOD}
In this section we provide an overview of the space-time POD method \citep{SchmidtSchmid2019JFM,FrameTowne2023PLoS}, focusing on its application to the analysis of flow transients. We seek the set of modes, $\vb*\phi(\vb*x,t)$, that optimally represent the second-order space-time statistics of the transient process, $\vb*q(\vb*x,t)$, over a finite time window, $t\in[t_0,t_0+\Delta T]$.
Here, $\vb*q$ represents the perturbation in \S\ref{sec:syncLES}. As $\vb*q$ is already a perturbation quantity, we will not again subtract the mean. But note that mean-flow deformations of the base state caused by the forcing will manifest as a steady component of $\vb*q$.
In general, $\vb*q$ may also be complex. To obtain the modes, we construct the weighted space-time inner product
\beq\label{eq:innerProd}
\expval{\vb*q_1(\vb*x,t),\vb*q_2(\vb*x,t)}_{\vb*x,t} = \int_{t_0}^{t_0+\Delta T}\!\!\!\!\!\int_\Omega \vb*q_2^*(\vb*x,t) \vb*W(\vb*x,t) \vb*q_1(\vb*x,t) \dd{\vb*x}\dd{t},
\eeq
where $\Omega$ is the domain of interest, $(\cdot)^*$ denotes the conjugate transpose, and $\vb*W(\vb*x,t)$ is a Hermitian positive definite weight tensor. The modes, $\vb*\phi(\vb*x,t)$, maximize the projection
\beq
\lambda = \frac{E\qty{|\expval{\vb*q(\vb*x,t),\vb*\phi(\vb*x,t)}_{\vb*x,t}|^2}}{\expval{\vb*\phi(\vb*x,t),\vb*\phi(\vb*x,t)}_{\vb*x,t}}. \label{eq:maximize}
\eeq
As equation~\eqref{eq:maximize} is the same quantity that is maximized by all variants of space-time POD \citep{SchmidtSchmid2019JFM,FrameTowne2023PLoS}, including SPOD \citep{TowneEtAl2018JFM}
and CS-SPOD \citep{HeidtColonius2024JFM}, the solutions, $\lambda_j$ and $\vb*\phi_j$, are given by the standard weighted Fredholm eigenvalue problem
\beq
\int_{t_0}^{t_0+\Delta T}\!\!\!\!\!\int_\Omega \vb*C(\vb*x,\vb*x',t,t')\vb*W(\vb*x',t')\vb*\phi(\vb*x',t')\dd{\vb*x'}\dd{t'} = \lambda\vb*\phi(\vb*x,t),
\eeq
where $\vb*C(\vb*x,\vb*x',t,t') = E\qty{\vb*q(\vb*x,t)\vb*q^*(\vb*x',t')}$ is the two-point space-time correlation tensor. Since $\vb*q$ is a fluctuation quantity, the eigenvalue, $\lambda_j$, measures the fluctuation energy corresponding to the mode, $\vb*\phi_j$, under the norm induced by the space-time inner product in equation~\eqref{eq:innerProd}. Properties analogous to those of more specialized forms of space-time POD also hold for this most general form. 
Unlike SPOD modes, space-time POD modes, $\vb*\phi(\vb*x,t)$, are time-varying. As such, the latter simultaneously provide both a statistical and a dynamical perspective on the evolution of coherent structures.

\section{Twin-rectangular jet flow}\label{sec:TRJflow}
\subsection{Numerical setup}\label{sec:numericalsetup}
\begin{table}
\begin{center}
\def~{\hphantom{0}}
\begin{tabular}{@{}lcccccccccc@{}}
    $M_j$ & $M_a$ & $Re_j$ & $p_t/p_\infty$ & $p_j/p_\infty$ & $T_t/T_\infty$ & $T_j/T_\infty$ & $N_{\mathrm{cv}}$ & $\dd t u_j/h$ \\
    1.5 & 1.25 & $1.07\times10^6$ & 3.671 & 1 & 1 & 0.69 & $77.0\times10^6$ & 0.00125 \\
\end{tabular}
\caption{Parameters of the synchronized simulations: jet Mach number, $M_j = u_j/c_j$, where $u_j$ and $c_j$ are the streamwise velocity and sound speed at the nozzle exit; acoustic Mach number, $M_a = u_j/c_\infty$, where $c_\infty$ is the ambient sound speed; Reynolds number, $Re_j=\rho_j u_j D_e/\mu_j$, where $\rho$ is the density, $\mu$ is the dynamic viscosity, $D_e=1.6h$ is the equivalent nozzle diameter, and $h$ is the nozzle height; nozzle pressure and temperature ratios, $p_t/p_\infty$ and $T_t/T_\infty$, where $(\cdot)_t$ refers to stagnation quantities; jet temperature ratio, $T_t/T_\infty$; grid size, $N_{\mathrm{cv}}$; and time step, $\dd t u_j/h$.}
\label{tab:LES_params}
\end{center}
\end{table}

The parameters of the LES closely follow the setup in our previous study of the forced supersonic twin-rectangular jet \citep{YeungSchmidt2023AIAA}, and are summarized in Table~\ref{tab:LES_params}. The simulations were carried out using the unstructured compressible solver `Charles' by Cadence.
The flow state has been non-dimensionalized by the jet exit conditions: density by $\rho_j$, velocities by $u_j$, temperature by $T_j$, and pressure by $\rho_ju_j^2$. Lengths are non-dimensionalized by $h$, and time by $h/u_j$. Dimensionless frequencies are expressed in terms of Strouhal numbers based on the equivalent nozzle diameter, $St = fD_e/u_j$.

The twin-rectangular nozzle geometry is described thoroughly by \citet{SamimyEtAl2023JFM}.
A cavity just upstream of each nozzle lip houses the plasma actuators, with six on each nozzle, consisting of three on the upper lip, and three on the lower lip.
For more details on the numerical setup, including its validation, we refer the reader to \citet{BresEtAl2021AIAA},
\citet{YeungEtAl2022AIAA}, and \citet{YeungSchmidt2023AIAA}.

To obtain multiple independent realizations for space-time POD, snapshots of the statistically stationary, natural jet are 
used as initial conditions for the transient simulations. Due to the present need to capture the intense disturbances created by the actuators, the grid used in the synchronized simulations has undergone slight refinement around the nozzle lip compared to the previous stationary simulation. The grid mismatch between the pre-existing and current computations causes a grid transient that we have confirmed has minimal effect on the perturbation. 
Based on estimates of the two-time autocorrelation, the initial conditions are spaced $(\Delta t)_\mathrm{IC}=100$ apart in time to guarantee their statistical independence. For simplicity and consistency, we set the length of the time window for space-time POD to be the same, $\Delta T=100$.

For this study, $N=17$ realizations of the transient were obtained for statistical analysis. Snapshots of the five primitive variables, $\rho$, $u$, $v$, $w$, and $T$, are extracted from the LES at intervals of $\Delta tu_j/h=0.25$, then interpolated from the unstructured grid onto a Cartesian grid. The variables are assembled in the state vector $\vb*q = [\rho,u,v,w,T]^\mathrm{T}$. Since the flow is compressible, we use the weight tensor
\beq\label{eq:chuWeight}
\vb*W(\vb*x,t) = \operatorname{diag}\qty(\qty[\frac{\tilde T(\vb*x,t)}{\gamma\tilde\rho(\vb*x,t) M^2_j}, \tilde\rho(\vb*x,t), \tilde\rho(\vb*x,t), \tilde\rho(\vb*x,t), \frac{\tilde\rho(\vb*x,t)}{\gamma(\gamma-1)\tilde T(\vb*x,t) M^2_j}]),
\eeq
where $\gamma$ is the adiabatic constant. The inner product in equation~\eqref{eq:innerProd} thus induces the compressible energy norm \citep{Chu1965ActaMech}.
The $\tilde{(\cdot)}$ notation refers to the ensemble mean estimated from all realizations.
Here, we have chosen the ensemble mean of the natural jet, since the natural jet is considered the base state in the framework of synchronized simulations.
To prevent the synchronized natural and forced jet flows from becoming entirely decorrelated due to turbulence, we restrict the spatiotemporal norm in equations~\eqref{eq:innerProd} and \eqref{eq:chuWeight} to $t\in[0,20]$ by setting the weight to zero outside this time window.
This window, $t\in[0,20]$, spans approximately 11 actuation periods and focuses on the initial transient.


\begin{figure}
    \centering
    \includegraphics{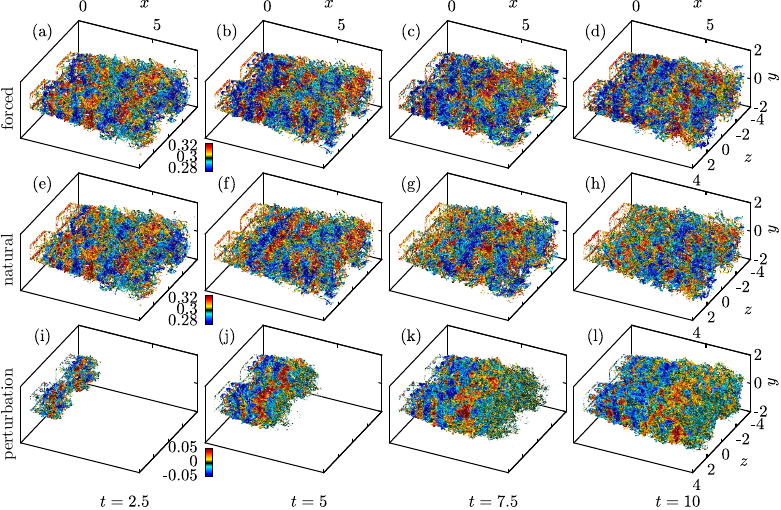}
    \caption{$Q$-criterion isocontours of $Q=5$, computed from the velocity perturbations at four time instances: (a) $t=2.5$; (b) $t=5$; (c) $t=7.5$; (d) $t=10$. The isocontours are colored by pressure: (a--d) $p_\mathrm{f}$; (e--h) $p_\mathrm{n}$; (i--l) $p_\mathrm{f}-p_\mathrm{n}$. Panels in each row share the same color bar.}
    \label{fig:qcriterion}
\end{figure}
Instantaneous snapshots of the forced jet, natural jet, and the perturbation are shown in figure \ref{fig:qcriterion} for one realization of the actuation transient. The snapshots are visualized using the $Q$ criterion \citep{HuntEtAl1988CTR} and display fully turbulent twin jet plumes. Pressure waves induced by the periodic forcing are visible in the region $x\lesssim5$ in both the forced jet and the perturbation. Because the perturbation is obtained as the difference between synchronized forced and natural simulations, the evolution of the transient loosely resembles the start-up of a jet exhausting into an initially quiescent medium. For avoidance of doubt, we reiterate that each simulation is initialized from a turbulent, statistically stationary state, not from a laminar base state or a uniform flow.

\subsection{Statistical symmetries}\label{sec:d2sym}
When symmetries are present in a flow, they should be exploited in modal decompositions, including space-time POD. Doing so reduces computational complexity, accelerates statistical convergence, and, perhaps most importantly, improves the interpretability of the results. The twin-rectangular jet nozzles are symmetric about the major and minor axes, $y=0$ and $z=0$, respectively. The geometry thus belongs in the dihedral group $D_2$. In the instantaneous flow, geometrical symmetries are broken by turbulence; however, they are imprinted on the turbulent statistics. Nevertheless, when the statistics are estimated from data, these symmetries will be imperfectly expressed, and must be enforced. In recent studies, we enforced $D_2$ symmetry on the SPOD analysis of the natural twin-rectangular jet \citep{YeungEtAl2022AIAA} and BMD analysis of the forced jet \citep{YeungSchmidt2023AIAA}. Here we extend the framework to space-time POD.

\begin{figure}
    \centering
    \includegraphics[width=\linewidth]{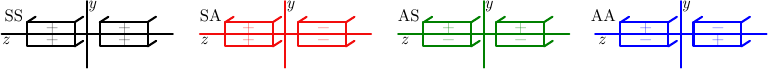}
    \caption{Illustration of $D_2$ symmetry decomposition \citep{YeungEtAl2022AIAA}.}
    \label{fig:symDecomp}
\end{figure}

Without loss of generality, the perturbations can be decomposed into four $D_2$ symmetry components (see e.g. \citet{SirovichPark1990POF}),
\beq
\vb*q^{(k)}(x,y,z,t) = \vb*q^{(k)}_\mathrm{SS}(x,y,z,t) + \vb*q^{(k)}_\mathrm{SA}(x,y,z,t) + \vb*q^{(k)}_\mathrm{AS}(x,y,z,t) + \vb*q^{(k)}_\mathrm{AA}(x,y,z,t),
\eeq
where the first and second letters in the subscripts denote symmetry (S) or antisymmetry (A) about the major and minor axes, respectively. The symmetry components are illustrated in figure \ref{fig:symDecomp}.
Following common practice, for each component we conduct an independent analysis using space-time POD. That is, we ignore nonlinear interactions between symmetry components \citep{YeungSchmidt2023AIAA}.

\subsection{Actuation}\label{sec:actuation}
The numerical modeling of the plasma actuators closely follows our recent work on the forced twin jet \citep{YeungSchmidt2023AIAA}, which was in turn an adaptation of the actuator model proposed by \citet{KimEtAl2009AIAA}. 
The actuation signal is a smoothed pulse wave with period $\tau$. Within each period, the actuator switches on at time $t_\mathrm{on}$, and off at time $t_\mathrm{off}$. The values for these parameters are estimated based on the available experimental measurements. 
The duty cycle, $(t_\mathrm{off}-t_\mathrm{on})/\tau\approx0.1\%$, is short. As a result, the actuation approximates an impulse train or Dirac comb, which is a periodic but non-harmonic forcing. All harmonics of the actuation frequency, $St_0$, are thus forced simultaneously.

To maximize control authority, the actuation must be consistent with the inherent symmetries of the flow, in this case $D_2$ symmetry. In the unforced case, the twin-rectangular jet is known to emit screech tones in the AS and AA symmetry components \citep{YeungEtAl2022AIAA}. For this work, we fire all actuators in-phase with each other, corresponding to the SS symmetry. The dynamics of the leading space-time POD modes of each symmetry will lend insight into the evolution of the jet flow from its preferred symmetries, AS and AA, to the forced symmetry, SS. In their experiments, \citet{SamimyEtAl2023JFM} demonstrated reduction in twin-rectangular jet noise using a forcing frequency of $St_0=0.9$. We therefore adopt the same forcing frequency.


\section{Transient dynamics and statistics}\label{sec:results}
\begin{figure}
    \centering
    \includegraphics{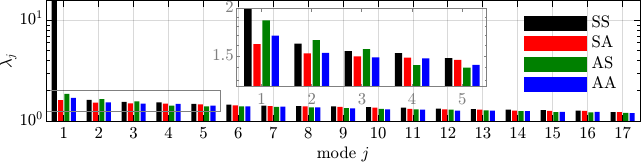}
    \caption{Space-time POD mode energy for each symmetry component. The inset zooms in on the leading five modes.}
    \label{fig:eigs}
\end{figure}
The space-time POD mode energies of each of the four $D_2$ symmetry components are shown in figure \ref{fig:eigs}. It is immediately apparent that the leading SS mode is extremely energetic. It represents $\lambda_1/\sum_j\lambda_j=41\%$ of the total energy in the SS component. For comparison, the leading mode in the SA, AS, and AA components each accounts for just 7--8\% of the total energy in the respective symmetry. The large separation between the leading and suboptimal eigenvalues in SS indicates low-rank behavior, which is commonly observed in SPOD and interpreted as a sign of dominant hydrodynamic instabilities or other physical mechanisms \citep{SchmidtEtAl2018JFM}. In this case, the rank separation in SS corresponds to the large perturbations to the natural flow that are directly linked to the SS forcing.

The SA, AS, and AA leading eigenvalues are all an order-of-magnitude less energetic than the SS leading mode. To study the differences between these low-energy modes, the inset in figure \ref{fig:eigs} zooms in on the first five eigenvalues for each symmetry, in the region $\lambda_j\in[1.25,2]$. Excluding the SS symmetry, the AS leading mode has the most energy. Among all four symmetries, AS also displays the fastest rank decay as the mode number, $j$, increases.

\begin{figure}
    \centering
    \includegraphics{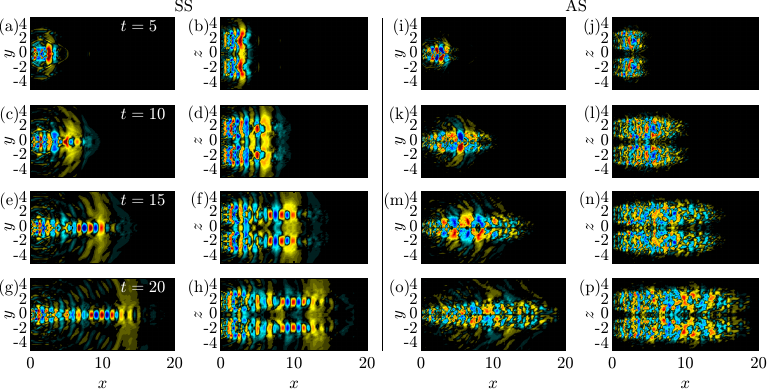}
    \caption{Pressure component of leading space-time POD mode: (a--h) SS; (i--p) AS symmetry components, visualized at $t=5$ (a,b,i,j), 10 (c,d,k,l), 15 (e,f,m,n), and 20 (g,h,o,p), on the $z=1.8$ (a,c,e,g,i,k,m,o) and $y=0.25$ (b,d,f,h,j,l,n,p) planes.  See supplementary movie 1 for an animation.}
    \label{fig:stpod_nsamp17_nt80}
\end{figure}
Figure \ref{fig:stpod_nsamp17_nt80} displays the pressure component of the leading space-time POD mode for the SS and AS symmetry components, at four representative time instances: $t=5,10,15,20$. The pressure component is reconstructed using the linearized ideal gas equation of state, $\phi_p=\tilde{\rho}\phi_T+\phi_\rho\tilde{T}$, where $\phi_\rho$ and $\phi_T$ are the density and temperature components, respectively, of the mode $\phi$, and $\tilde{(\cdot)}$ represents the ensemble mean of the natural jet. For the SS symmetry in figure \ref{fig:stpod_nsamp17_nt80}(a--h), we observe an acoustic wave front that is created by the initial pulse of the actuator, corresponding to the response of the natural jet to the impulsive forcing. This wave front propagates downstream at the speed of sound. Subsequent actuation cycles also generate aft-angle acoustic waves at the forcing frequency, $St_0$, but with significantly lower amplitudes than the impulse response. 

\begin{figure}
    \centering
    \includegraphics{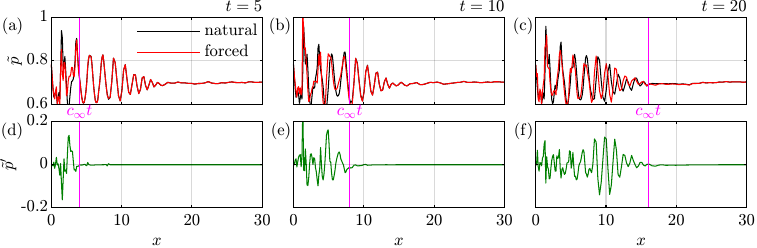}
    \caption{Pressure along the nozzle centerline, $y=0$ and $z=1.8$, for the SS component: (a--c) ensemble mean of the natural and forced transients; (d--f) difference between the respective ensemble mean. Three time instances are displayed: $t=5$ (a,d), $10$ (b,e), and $20$ (c,f). The estimated location of the initial acoustic wave, $x=c_\infty t$, is marked by the magenta line.  See supplementary movie 2 for an animation.}
    \label{fig:p_centerline_SS}
\end{figure}
Behind the impulsive wave front, in the downstream region $x\gtrsim5$, modes are visible along the centerline of each jet. Close examination of the four time instances reveals that the nodes and antinodes of these modes are stationary in space. These modes are not hydrodynamic in origin. Instead, they result from forcing-induced deformations of the mean flow, specifically the shock cell structure. Figure \ref{fig:p_centerline_SS} compares the ensemble mean of the transients of the natural and forced jets, as well as their difference, along the centerline of one nozzle, at $y=0$ and $z=1.8$, for the SS symmetry component. Also shown is the location of the initial acoustic wave front, estimated from the ambient sound speed, $c_\infty$. The acoustic wave clearly demarcates the region as yet unaffected by the forcing, $x>c_\infty t$, from the region that has felt its influence, $x<c_\infty t$. For $x>c_\infty t$, the natural and forced mean flows overlap. For $x<c_\infty t$, the two deviate from each other. In particular, the forcing causes the shock cells to shift further upstream, perhaps entailing a slight contraction of the potential core length. The steady phase difference between the shocks of the natural and forced jets explains the wave-like pattern along the centerline of the SS space-time POD mode.


Returning to the POD modes in figure \ref{fig:stpod_nsamp17_nt80}(i--p), the AS symmetry component, which is known to manifest jet screech in the natural twin-rectangular jet \citep{YeungEtAl2022AIAA}, does not exhibit the upstream-propagating hydrodynamic and acoustic waves that are required to close the screech feedback loop \citep{Edgington-MitchellEtAl2022JFM}. This indicates the screech feedback mechanism is present in both jets. Moreover, the waves involved in screech in both jets remain in phase with each other, so that they vanish in the perturbation, i.e., the difference between the two transients. It is thus unlikely that the present actuation strategy can achieve significant screech mitigation. 

Rather, the main effect of the forcing on the AS component is the impulse response from the initial actuation cycle. In contrast to the SS mode in figure \ref{fig:stpod_nsamp17_nt80}(a,b), small-scale turbulent structures are generated almost instantaneously in the AS mode in \ref{fig:stpod_nsamp17_nt80}(i,j). Superimposed on the background turbulence is a dispersive wavepacket that emerges in \ref{fig:stpod_nsamp17_nt80}(i,j), then increases its wavelength from \ref{fig:stpod_nsamp17_nt80}(i,j) through \ref{fig:stpod_nsamp17_nt80}(n), before disappearing in \ref{fig:stpod_nsamp17_nt80}(o,p). Subsequent actuation cycles do not create such a wavepacket.

\section{Discussion and summary}\label{sec:discussion}
When a statistically stationary turbulent flow is subjected to periodic forcing, it first undergoes a transient state before eventually attaining statistical cyclostationarity. During its transient phase, the statistics of the flow are time-varying and aperiodic. Consequently, 
we use the most general form of space-time POD to study actuation transients. We demonstrate this approach on the transients of a plasma-controlled supersonic twin-rectangular jet, periodically and symmetrically forced at $St=0.9$. To help isolate the perturbations that arise due to the forcing from the background turbulence, we conduct synchronized simulations of the natural and forced jets, then perform space-time POD on a statistical ensemble of their difference. The difference, or perturbation, visually resembles a jet starting up from a uniform initial condition. Because of this, we expect space-time POD to also be well-suited to the analysis of true start-up processes.

The space-time POD eigenvalues reveal low-rank behavior in the SS symmetry component, whose perturbation energy is well-captured by the leading mode. The remaining three symmetry components---SA, AS, and AA---exhibit slow rank decay. The SS and AS leading modes show that the initial pulse of the actuator has the greatest effect, generating a high-amplitude acoustic wave in the SS mode, and a wavepacket in the AS mode that briefly emerges before dissolving into turbulence. The SS mode highlights the mean flow deformation transient occurring near the nozzle centerline, specifically, a reduction in shock spacing due to the forcing. No significant effect on screech is observed.


The modes in figure \ref{fig:stpod_nsamp17_nt80} highlight an obvious limitation when applying space-time POD to transients: a very large number of realizations may be required for statistical convergence. When an appropriate  spatial (circular, reflectional, etc.) symmetry or temporal (periodic, stationary, cyclostationary, etc.) symmetry is known \textit{a priori} and imposed on the decomposition, leveraging the symmetry acts as a powerful filter for the statistics. In the case of a statistically stationary round jet, for instance, the solutions to space-time POD are known to be sinusoidal in both azimuth and time. In this work, with the exception of $D_2$ symmetry, no solution symmetry is imposed, and the price we pay for generality is slow statistical convergence.

To improve convergence, the strategy we chose is to perform synchronized natural and forced simulations, followed by space-time POD of the perturbation. The perturbation, however, includes the difference between the mean flows of the natural and forced transients. This differs from the convention in POD, in which a perturbation arises from removing the mean, and the modes solely represent variance.
In the present study, inclusion of the ensemble mean is motivated by the rapid rank decay of the space-time POD energy spectrum (see figure \ref{fig:eigs}). The small suboptimal eigenvalues demonstrate that the transients exhibit little variance about the ensemble mean, and that the transient dynamics are nearly entirely determined by the ensemble mean. This observation has important implications for physical modeling, as a time-varying mean flow can be economically predicted using tools such as unsteady Reynolds-averaged Navier-Stokes (URANS) solvers. It is worth pointing out that among previous efforts to apply space-time POD to transients, it is in fact common not to subtract the ensemble mean (see e.g. \citet{SchmidtSchmid2019JFM}, \citet{BorraSaxton-Fox2022AIAAJ}, and others),
a sign that the mean often contains the physics of interest. For transient flows that exhibit significant variance about the mean, a mean-subtracted space-time POD is more suitable \citep{KernEtAl2024JFM}. Whether general trends exist that lead to transient flows with high or low variance is a topic we plan to assess in future work.

\backsection[Funding]{We gratefully acknowledge support from Office of Naval Research grant N00014-23-1-2457, under the supervision of Dr. Steve Martens. 
}

\backsection[Declaration of interests]{The authors report no conflict of interest.}




\bibliographystyle{jfm}
\bibliography{ref}


\end{document}